\documentclass[12pt]{article}
\pdfoutput=1

\usepackage[bulletsep]{collref}
\usepackage{amssymb,graphicx}
\usepackage[intlimits]{amsmath}
\usepackage{bbm}
\usepackage[small]{subfigure}

\usepackage{MnSymbol}

\usepackage{ytableau}

\makeatletter \@addtoreset{equation}{section} \makeatother

\makeatletter
\let\old@startsection=\@startsection
\let\oldl@section=\l@section
\renewcommand{\@startsection}[6]{\old@startsection{#1}{#2}{#3}{#4}{#5}{#6\mathversion{bold}}}
\renewcommand{\l@section}[2]{\oldl@section{\mathversion{bold}#1}{#2}}
\makeatother

\makeatletter
\let\old@makecaption=\@makecaption
\def\@makecaption{\small\old@makecaption}
\makeatother

\renewcommand{\leq}{\leqslant}

\newcommand{\F}{\mathcal{F}}

\newcommand{\f}{\mathtt{f}}

\begin{document}

\begin{flushright}\footnotesize
\texttt{NORDITA-2015-14} \\
\texttt{UUITP-03/15}
\vspace{0.6cm}
\end{flushright}

\renewcommand{\thefootnote}{\fnsymbol{footnote}}
\setcounter{footnote}{0}

\begin{center}
{\Large\textbf{\mathversion{bold} 
Higher Rank
Wilson Loops 
\\ 
in $N=2^*$ Super-Yang-Mills Theory
}
\par}

\vspace{0.8cm}

\textrm{Xinyi~Chen-Lin and
Konstantin~Zarembo\footnote{Also at ITEP, Moscow, Russia}}
\vspace{4mm}

\textit{Nordita, KTH Royal Institute of Technology and Stockholm University,
Roslagstullsbacken 23, SE-106 91 Stockholm, Sweden}\\
\textit{Department of Physics and Astronomy, Uppsala University\\
SE-751 08 Uppsala, Sweden}\\
\vspace{0.2cm}
\texttt{xinyic@nordita.org, zarembo@nordita.org}

\vspace{3mm}


\par\vspace{1cm}

\textbf{Abstract} \vspace{3mm}

\begin{minipage}{13cm}
The $\mathcal{N}=2^*$ Super-Yang-Mills theory (SYM*)  undergoes an infinite sequence of large-$N$ quantum phase transitions. We compute expectation values of Wilson loops in $k$-symmetric and antisymmetric representations of the $SU(N)$ gauge group in this theory and show that the same phenomenon that causes the phase transitions  at finite coupling leads to  a non-analytic dependence of Wilson loops on $k/N$ when the coupling is strictly infinite, thus making the higher-representation Wilson loops ideal holographic probes of the non-trivial phase structure of SYM*.
\end{minipage}

\end{center}

\vspace{0.5cm}

\renewcommand{\thefootnote}{\arabic{footnote}}
\setcounter{footnote}{0}

\section{Introduction}

The path integral of any $\mathcal{N}=2$ gauge theory on $S^4$ can be calculated \cite{Pestun:2007rz} by means of the supersymmetric localization \cite{Witten:1988ze}. 
Localization reduces the path integral to a matrix model giving us direct access to strong coupling, and in particular to the regime of interest for holography, when $N$ is infinite and the 't~Hooft coupling $\lambda =g_{\rm YM}^2N$ is large. 

The SYM* theory is particularly well-suited for this purpose, since its holographic dual is explicitly known \cite{Pilch:2000ue}, 
while the $\mathcal{N}=2^*$ localization matrix model can be solved at strong coupling \cite{Buchel:2013id,Chen:2014vka,Zarembo:2014ooa} 
by more or less standard methods of random matrix theory \cite{Brezin:1977sv}. 
Various predictions of holography can then be confronted with {\it ab initio} evaluation of the field-theory path integral \cite{Buchel:2013id,Bobev:2013cja}. 
Perfect agreement found so far has left, nevertheless, one feature unexplained from the holographic perspective. 
Localization predicts an infinite sequence of quantum phase transitions that occur at certain critical values of the 't~Hooft coupling in the large-$N$ limit \cite{Russo:2013qaa,Russo:2013kea}. Large-$N$ phase transitions are of course very common in matrix models \cite{Gross:1980he,Wadia:2012fr}, 
but here the real critical behavior occurs such that correlation length goes to infinity. The ensuing phase transitions also have finite-$N$ counterparts \cite{Russo:2014nka}.
Similar transitions occur in other localization path integrals \cite{Russo:2013kea,Barranco:2014tla,Anderson:2014hxa,Minahan:2014hwa}, 
and it would be very interesting to understand their holographic origin.

Unfortunately, tuning the 't~Hooft coupling to some given critical value is very difficult in holography. 
Well-established holographic methods work at strictly infinite coupling, leaving us little hope to observe the transitions as discontinuities of correlation functions in $\lambda $.
However, in the SYM* the $n$-th critical coupling grows with $n$  as $\lambda _c^{(n)}\simeq \pi ^2n^2$ \cite{Russo:2013kea}, and most of the phase transitions do happen when the coupling is very large. Certain observables may then have criticality-induced non-analyticities even when the coupling is strictly infinite. Identifying such observables is the goal of this paper.

The phase transitions are driven by nearly massless states that give rise to singularities in the eigenvalue density of the localization matrix model. 
The eigenvalue density can be directly probed by D-branes in the dual supergravity background \cite{Buchel:2000cn}. 
The classical D-branes see  the coarse-grained density, in which the resonances are smoothened out \cite{Russo:2013qaa}, 
but D-brane quantum fluctuations may be sensitive to the non-trivial phase structure of SYM*.

Another type of observables calculable by localization are Wilson loops. They might appear useless for detecting phase transitions at first sight. There are two scales in the problem, $M$ and $1/R$, where  $M$ is the mass scale of the SYM* theory and $R$ is the radius of compactification on $S^4$. For the most part of this paper we regard the radius of $S^4$ as an IR regulator and thus assume that $MR\gg 1$. At strong coupling a new scale $\mu \sim \sqrt{\lambda }M\gg M$ emerges. Wilson loops in the fundamental representation probe the smallest scales of order $1/R$, the D-branes probe the largest scale of order $\mu $, while resonances that cause critical behavior occur at the intermediate scale of order $M$. In this paper we consider Wilson loops in higher representations of the gauge group, and show that by varying the size of representation it is possible to scan the whole spectrum of scales, including the resonance region.

Wilson loops of rank $k\sim N$ are dual to D-brane configurations carrying the fundamental string charge \cite{Drukker:2005kx,Yamaguchi:2006tq,Gomis:2006sb,RodriguezGomez:2006zz,Gomis:2006im,Yamaguchi:2007ps,Fiol:2014vqa}, and have been extensively studied in the $\mathcal{N}=4$ super-Yang-Mills theory \cite{Yamaguchi:2006tq,Okuyama:2006jc,Hartnoll:2006is,Yamaguchi:2007ps,Kawamoto:2008gp,Fiol:2013hna,Buchbinder:2014nia,Faraggi:2014tna}, for which the matrix model is just Gaussian \cite{Erickson:2000af,Drukker:2000rr}. As far as $\mathcal{N}=2$ theories are concerned, higher-rank Wilson loops have been  calculated for the $\mathcal{N}=2$ superconformal QCD \cite{Fraser:2011qa},   whose solution at strong coupling is known \cite{Passerini:2011fe} but is very different from that of SYM*. 

\section{Wilson loops}

The SYM* theory has the same field content as $\mathcal{N}=4$ SYM -- gauge fields, six adjoint scalars and four Majorana fermions, but its Lagrangian includes explicit mass terms for half of the fields (four scalars and their $\mathcal{N}=2$ superpartners). We denote this common mass by $M$.

The Wilson loop in representation $\mathcal{R}$ is defined as
\begin{equation}\label{Wilsoncontour}
 W_\mathcal{R}(C)=\left\langle \mathop{\mathrm{tr}}\!{}_\mathcal{R}\mathop{\mathrm{P}}\exp
 \left[
 \oint_C ds\,\left(
 i\dot{x}^\mu A_\mu +|\dot{x}|\Phi 
 \right)
 \right]
 \right\rangle,
\end{equation}
where $\Phi $ is a scalar from the vector multiplet, coupling to which makes the Wilson loop locally supersymmetric and well-defined in  the UV. 

Localization computes the Wilson loop for the equatorial contour of $S^4$ \cite{Pestun:2007rz}. The partition function on the four-sphere of radius $R$ localizes to an eigenvalue integral:
\begin{equation}\label{Zed}
 Z=\int_{}^{}d^{N-1}a\,\prod_{i<j}^{}\mathcal{Z}_{\rm 1-loop}\left(a_i-a_j\right)
 \,{\rm e}\,^{-\frac{8\pi ^2NR^2}{\lambda }\,\sum_{i}^{}a_i^2}.
\end{equation}
The integration variables are the Coulomb moduli that parameterize space-time independent zero modes of the scalar that enters the loop operator:
\begin{equation}\label{vacuum-field}
\Phi_0=\mathop{\mathrm{diag}}\left(a_1,\ldots ,a_N\right).
\end{equation}
The non-zero modes of $\Phi $, as well as all other degrees of freedom have been integrated out, leaving behind a one-loop contribution \cite{Pestun:2007rz}:
\begin{equation}\label{Z1loop}
 \mathcal{Z}_{\rm 1-loop}(x)=\frac{x^2H^2(x)}{H(x+M)H(x-M)}\,,\qquad H(u)=\prod_{n=1}^\infty \left(1+\frac{R^2u^2}{n^2}\right)^n \,{\rm e}\,^{-\frac{R^2u^2}{n}} .
\end{equation}

The Wilson loop expectation value is obtained by simply replacing quantum fields in (\ref{Wilsoncontour}) by their values on the localization locus and subsequently averaging over the Coulomb moduli:
\begin{equation}\label{localized-loop}
 W_\mathcal{R}(C)=\left\langle \mathop{\mathrm{tr}}\!{}_\mathcal{R}
 \,{\rm e}\,^{L\Phi _0}
 \right\rangle.
\end{equation}
We denote by $L$ the length of the contour $C$. The average is now defined by the eigenvalue partition function (\ref{Zed}).

When $C$ is the big circle on $S^4$, this result is exact (up to instanton corrections neglected here because of the large-$N$ suppression),  in which case \begin{equation}
 L=2\pi R.
\end{equation}
Localization, strictly speaking, is not applicable to other contours, but the leading exponential behavior for asymptotically large Wilson loops should be universal and largely insensitive to the contour's shape. We thus expect that our results apply to any sufficiently big contour, including contours on $\mathbbm{R}^4$, as long as $ML\gg 1$. This assertion is based on universality  and has less rigorous grounds compared to localization, but has been checked holographically for Wilson loops in the fundamental representation \cite{Buchel:2013id}.

We concentrate on rank-$k$ symmetric or antisymmetric representations:
\begin{equation}
\ytableausetup{mathmode, smalltableaux, centertableaux}
 \mathcal{R}_k^+=
 \overbrace{
 \begin{ytableau}
 ~ & ~ &~ & ~ &  \\
 \end{ytableau}}^{k}
 \qquad 
 \mathcal{R}_k^-=
 \left.
 \begin{ytableau}
 ~ \\ ~ \\ ~ \\ ~ \\ ~ \\
 \end{ytableau}
 \right\}k
\end{equation}
and will eventually consider the scaling limit $N\rightarrow \infty $, $k\rightarrow \infty $, in which the ratio
\begin{equation}\label{f-def}
 f=\frac{k}{N}
\end{equation}
is kept fixed.
Such Wilson loops are dual to macroscopic D3 or D5 branes, depending on whether representation is symmetric or antisymmetric \cite{Drukker:2005kx,Yamaguchi:2006tq,Gomis:2006sb,Gomis:2006im,Yamaguchi:2007ps}, that carry $k$ units of electric flux on their world-volume. On the field-theory side, the 
$k$-symmetric/antisymmetric Wilson loops are described by statistical mechanics of free bosons/fermions, for which the matrix eigenvalues play the r\^ole of energy levels \cite{Hartnoll:2006is}. Let us review how this statistical interpretation arises in the matrix model.

The $k$-symmetric or antisymmetric characters are conveniently packaged into  generating functions:
\begin{equation}
 \chi ^\pm (\nu ,E)=\sum_{k=0}^{\infty }\,{\rm e}\,^{-k\nu }\mathop{\mathrm{tr}}\!{}_{\mathcal{R}_k^\pm}\,{\rm e}\,^{E} .
\end{equation}
 For the antisymmetric representations, the sum terminates at $k=N$. An $SU(N)$ character  is a class function, in other words, a symmetric function of the eigenvalues $\varepsilon _n$ of the matrix $E$. When expressed in terms of eigenvalues, the generating functions can be explicitly written as Bose and Fermi distribution:
\begin{equation}
 \chi ^\pm (\nu ,E)=\prod_{n=1}^{N}\left(1\mp \,{\rm e}\,^{\varepsilon_n-\nu  } \right)^{\mp 1},
\end{equation}
 where (minus) the eigenvalue plays the r\^ole of the energy level and $-\nu $ plays the r\^ole of chemical potential. 
 
Using this representation, we find for the Wilson loop (\ref{localized-loop}) in the $k$-symmetric/antisymmetric representation:
\begin{equation}\label{a-average}
 W_{\mathcal{R}_k^\pm}\equiv W_k^\pm=L\left\langle 
 \int_{C-i\pi }^{C+i\pi }\frac{d\nu }{2\pi i}\,\,\,{\rm e}\,^{kL\nu }\prod_{j}^{}\left[1\mp\,{\rm e}\,^{L(a_j-\nu) }\right]^{\mp 1}
 \right\rangle.
\end{equation}
For symmetric representations, the contour of integration should be chosen such that $C>a_j$ for any $j$. 
For antisymmetric representations, $C$ is arbitrary.
We have rescaled $\nu $ by $L$ for future convenience.
 
 The vacuum distribution of eigenvalues $a_j$ is characterized by the eigenvalue density:
\begin{equation}
 \rho (x)=\left\langle \frac{1}{N}\sum_{j}^{}\delta \left(x-a_j\right)\right\rangle.
\end{equation}
At large-$N$ the density  does not fluctuate, because the action in (\ref{Zed}) is $O(N^2)$ while there are only $N$ variables of integration. The weight in the integral representation  for the rank-$k$ Wilson loop (\ref{a-average}) is exponentially large, but the exponent is $O(N)$ rather than $O(N^2)$, and the insertion of the Wilson loop does not affect the saddle point distribution of eigenvalues to the leading order in $1/N$. The statistical average in (\ref{a-average}) is then replaced by the average with respect to the fixed ensemble determined by the master field $\rho (x)$\footnote{This is not quite true for the symmetric representations in certain range of parameters, as we shall discuss in  later.}:
\begin{equation}\label{contour-int}
 W_k^\pm\simeq L\int_{C-i\pi }^{C+i\pi }\frac{d\nu }{2\pi i}\,\,
 \,{\rm e}\,^{NL\mathcal{F}^\pm(\nu )},
\end{equation}
where
\begin{equation}\label{FreeE}
 \F^\pm(\nu )=f\nu\mp\frac{1}{L}\int_{-\mu }^{\mu }dx\,\rho (x)\ln\left(1\mp\,{\rm e}\,^{L(x-\nu )}\right),
\end{equation}
and $f$, defined in (\ref{f-def}), is assumed to be of order one in the large-$N$ limit.

The contour integral in (\ref{contour-int}) is of the saddle-point type, and to the leading order in $1/N$
we find:
\begin{equation}\label{logW}
 \ln W_k^\pm = NL \F^\pm(\nu )+O\left(N^0\right),
\end{equation}
where $\nu $ is determined by minimizing  the free energy:
\begin{equation}\label{cp<->dens}
 \int_{-\mu }^{\mu }\frac{dx\,\rho (x)}{\,{\rm e}\,^{L(\nu -x)}\mp 1}=f.
\end{equation}
This is the standard relation between canonical and grand canonical ensembles in statistical mechanics, with $-\nu $ playing the r\^ole of  chemical potential, $f$  the density of particles and  $\rho (x)$ the level density. As expected,  the exponent in the expectation value of the Wilson loop is proportional to the length of the contour. 
To compute the coefficient of proportionality, we first need to solve for the eigenvalue density of the matrix model (\ref{Zed}), then find $\nu $  from (\ref{cp<->dens}) and substitute the result into (\ref{FreeE}).  

The  eigenvalue density for the localization matrix model is not known in general, but at large $R$ and large $\lambda $,
a systematic perturbative solution can be constructed. The leading-order strong coupling solution was known for long time from the D-brane probe analysis of the dual supergravity background  \cite{Buchel:2000cn} and can be easily reproduced from the saddle-point equations of the matrix model \cite{Buchel:2013id}. It has the form of the Wigner distribution:
\begin{equation}\label{rho_inf}
 \rho _\infty (x)=\frac{2}{\pi \mu ^2}\,\sqrt{\mu ^2-x^2},
\end{equation}
with the width proportional to the square root of the 't~Hooft coupling\footnote{This results holds in the infinite volume. The finite-$R$ expression for the partition function on $S^4$ is obtained by  replacing $M$ with $\sqrt{M^2+1/R^2}$  \cite{Buchel:2013id}.}:
\begin{equation}\label{mu}
 \mu =\frac{\sqrt{\lambda }M}{2\pi }\,.
\end{equation}

The leading-order is way too simple to describe the critical behavior of the model, which arises because of the complicated short-distance structure of the density, that at the leading order gets averaged over. Interestingly, the first strong-coupling correction fully reveals the short-distance singularities that are responsible for quantum phase transitions at finite $\lambda $ \cite{Chen:2014vka,Zarembo:2014ooa}.

\begin{figure}[t]
\begin{center}
 \centerline{\includegraphics[width=11cm]{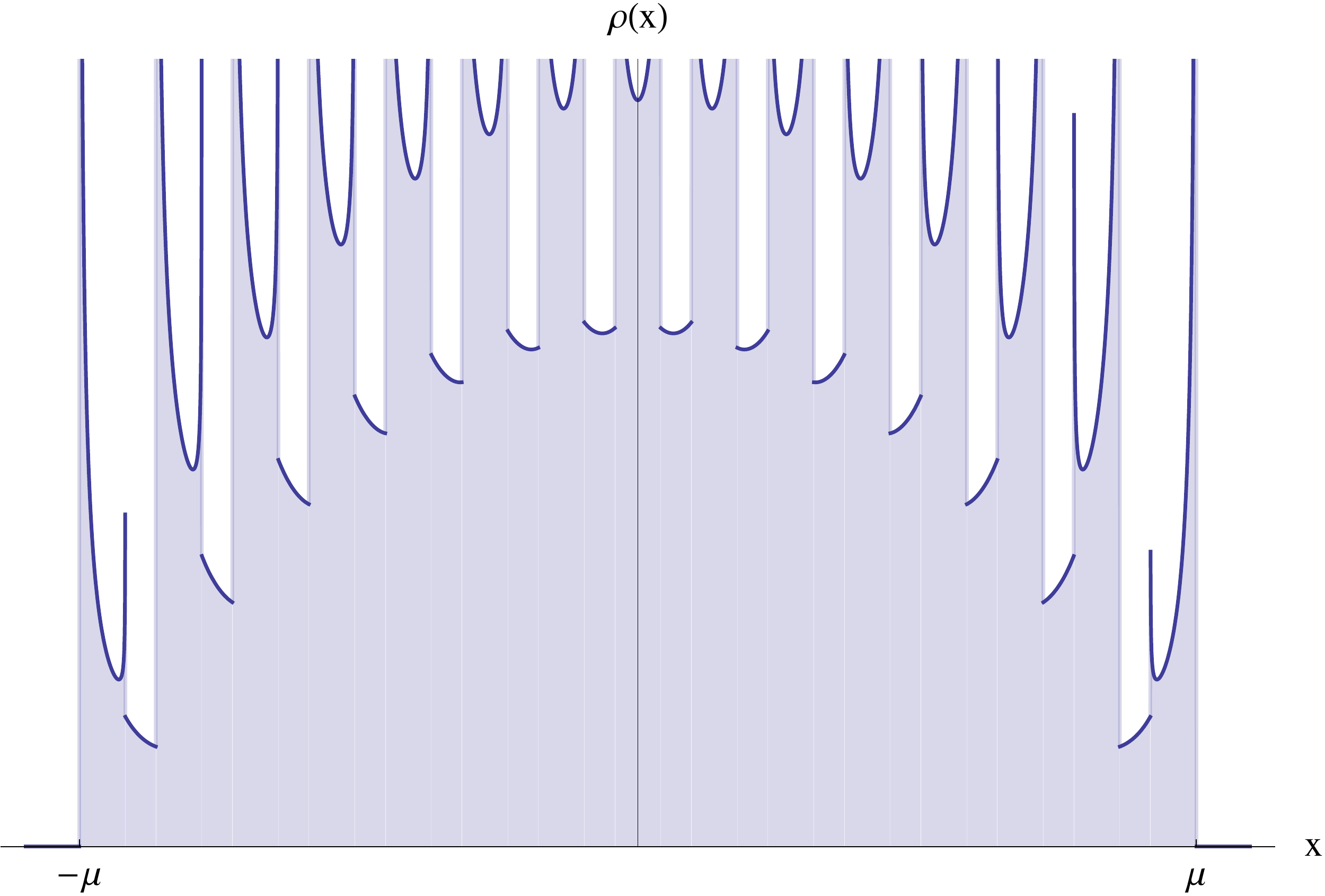}}
\caption{\label{fig-density}\small The eigenvalue density for $\mu =7.3M$.}
\end{center}
\end{figure}

Although we expect from holography that the strong-coupling expansion goes in powers of $1/\sqrt{\lambda }$, the first correction to the density appears at the relative order $(M/\mu)^{1/2} \sim \lambda ^{-1/4}$. The correction has a very irregular, spiky structure (fig.~\ref{fig-density}) \cite{Chen:2014vka,Zarembo:2014ooa}:
\begin{eqnarray}\label{bulk-density}
 \rho  (x)&\simeq& \frac{2}{\pi \mu ^2}\,\sqrt{\mu ^2-x^2}
\nonumber \\
&&
 +\frac{1}{\pi }\sqrt{\frac{M}{2\mu ^5}}\left[
 \left(\mu -x\right)\zeta \left(\frac{1}{2}\,,\left\{\frac{\mu +x}{M}\right\}\right)
 +\left(\mu +x\right)\zeta \left(\frac{1}{2}\,,\left\{\frac{\mu -x}{M}\right\}\right)
 \right],
\end{eqnarray}
where $\left\{\cdot\right\}$
denotes the fractional part. The zeta-function $\zeta (1/2, z)$  has a  $z^{-1/2}$
singularity at zero, and this produces an array of infinite spikes at resonance points $x=\pm \mu \mp nM$. 

The physical origin of the spiky structure in the density is due to the resonance on massless hypermultiplets.
The number of resonances that fit into the interval $(-\mu ,\mu )$ depends on the ratio $2\mu /M$ 
and each time $2\mu $ crosses an integer multiple of $M$, a new pair of resonances appears, 
leading to a fourth-order quantum phase transition \cite{Russo:2013qaa,Russo:2013kea}. 
The transitions are sharp only in infinite volume, at finite $R$ the spikes are rounded up and the transitions become  smooth crossovers. 
The expression above assumes that the infinite volume limit is taken prior to the strong-coupling limit, in the sense that $MR\gg \sqrt{\lambda }$. If the strong-coupling limit is taken first, the spikes get damped, albeit  at a very slow rate, and completely disappear distance $M^2R\sim \mu MR/\sqrt{\lambda }$ away from the endpoints \cite{Chen:2014vka}.

\begin{figure}[t]
\begin{center}
 \centerline{\includegraphics[width=10cm]{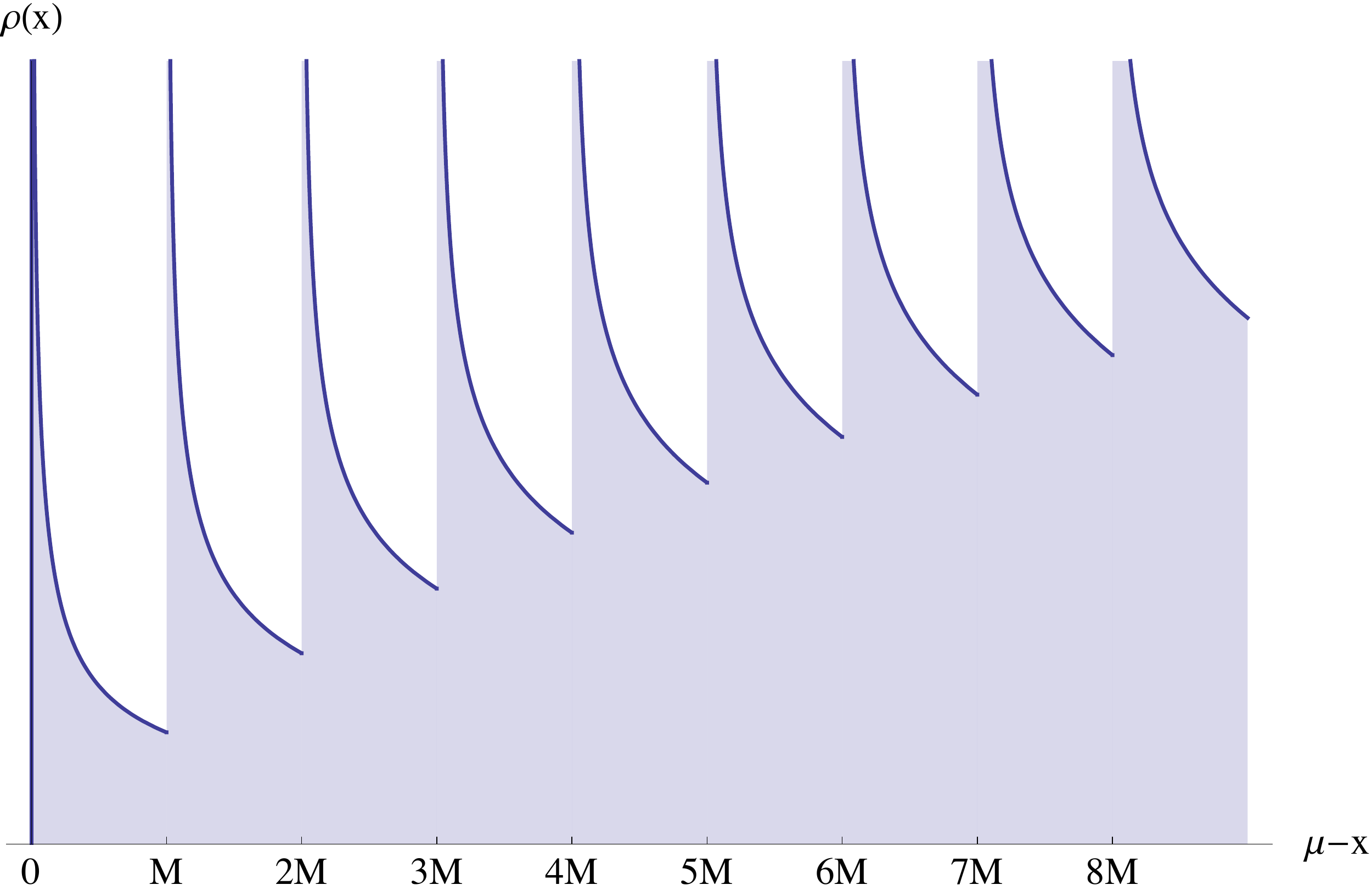}}
\caption{\label{fig-density-edge}\small The density near the endpoint of the eigenvalue distribution.}
\end{center}
\end{figure}

In the bulk of the eigenvalue distribution, the comb-like structure of resonances is a small correction, which moreover disappears upon averaging over a sufficiently wide interval because
\begin{equation}
 \int_{0}^{1}d\xi \,\zeta \left(\frac{1}{2}\,,\xi \right)=0.
\end{equation}
However, near the endpoints the spikes are no longer suppressed compared to the averaged density.  Indeed the two terms in (\ref{bulk-density}) become comparable for $\mu -x\sim M$. The whole expression is actually not applicable near the endpoints.
The edge behavior of the eigenvalue density is described instead by \cite{Chen:2014vka}
\begin{equation}\label{edge-d}
\rho (x)
 =\frac{1}{\pi }\sqrt{\frac{2M}{\mu ^3}}\sum_{k=0}^{\left[\frac{\mu -x}{M}\right]}\frac{1}{\sqrt{\left\{\frac{\mu -x}{M}\right\}+k}}\qquad \left(\mu -x\sim 1\right),
\end{equation}
where $\left[\cdot\right]$ denotes the integer part. This function, shown in fig.~\ref{fig-density-edge},  smoothly matches with (\ref{bulk-density}) at $\mu -x\gg M$.

\begin{figure}[t]
\begin{center}
 \centerline{\includegraphics[width=10cm]{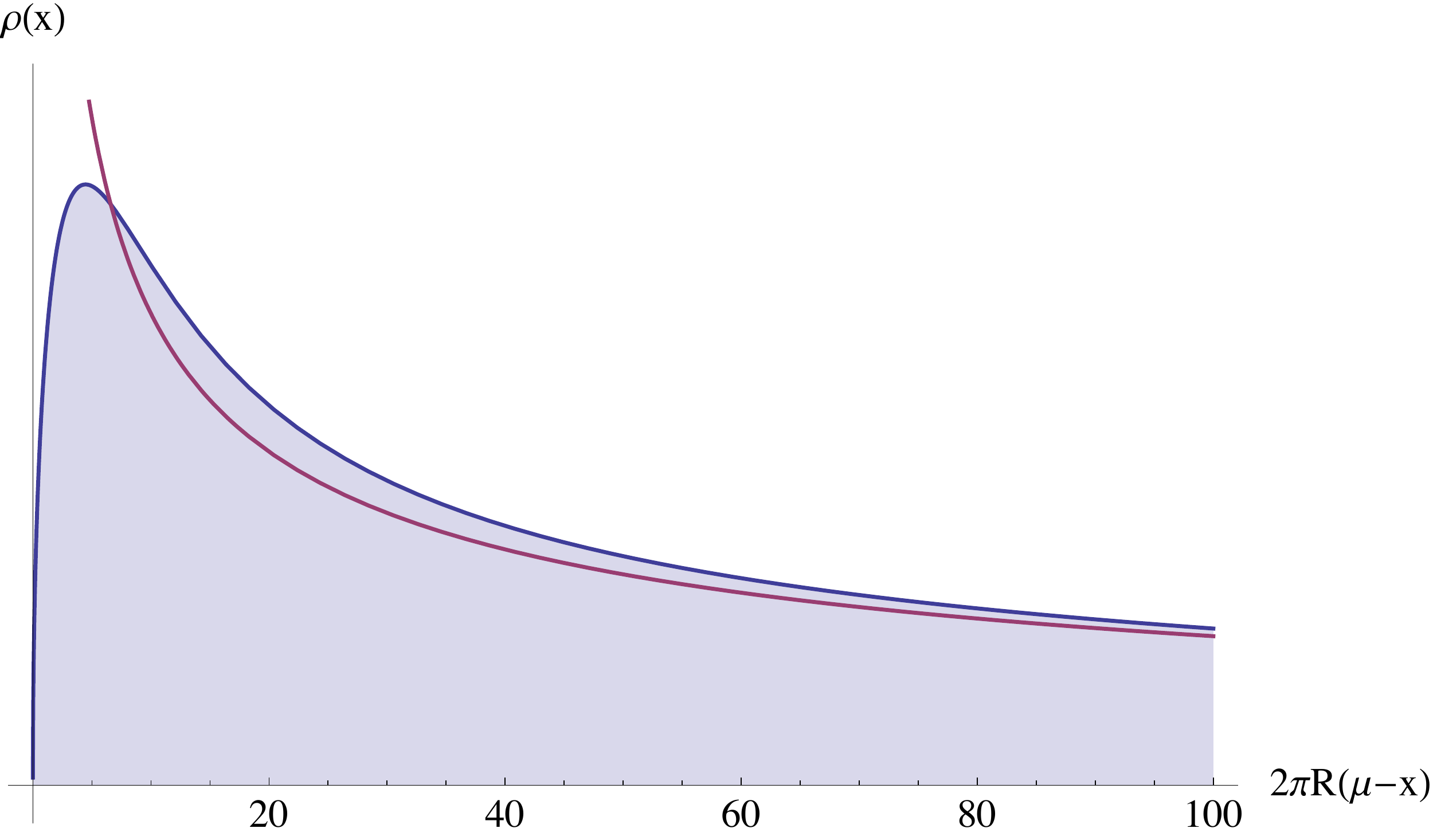}}
\caption{\label{fig-dens-1-peak}\small The finite-$R$ resolution of the first peak. The density (\ref{edge-d}) with the peak unresolved is shown in the purple line.}
\end{center}
\end{figure}

Extremely close to the endpoint, at $\mu -x\sim 1/R$, the structure of the  peak starts to be resolved. The density has the following form in this regime \cite{Chen:2014vka}:
\begin{equation}\label{densityPeak1}
 \rho (x)=\frac{2M}{\pi ^3}\sqrt{\frac{R}{\mu ^3}}\int_{0}^{\infty }
 \frac{d\kappa }{\sqrt{\kappa }}\,\,\cos2\pi \kappa \sin^2\pi \kappa \,\Gamma ^2(\kappa )
 \,{\rm e}\,^{-2\kappa \left(\ln\kappa -1\right)-2\pi R\left(\mu -x\right)\kappa },
\end{equation}
and is shown in fig.~\ref{fig-dens-1-peak}. This regime exists only for the theory defined on $S^4$. 
The density near the edgepoints is known even without assuming that $MR\gg 1$ \cite{Chen:2014vka}, but the general expression is substantially more complicated. 

With the density at hand, we can now compute the Wilson loop expectation value from (\ref{cp<->dens}), (\ref{logW}), (\ref{FreeE}). 
Depending on the parameter $f$, the main contribution to the integrals in (\ref{cp<->dens}), (\ref{FreeE}) will come from different range of $x$'s. We may identify three distinct regimes:
\begin{itemize}
 \item {\bf Bulk regime:} $\mu -x\sim O(\mu )$. In this case, the average density  (\ref{rho_inf}) is a good approximation. The spikes constitute a $\lambda ^{-1/4}$ correction as clear from (\ref{bulk-density}). 
 \item {\bf Endpoint regime:} $\mu -x\sim O(M)$. The spiky structure of the density then appears in the leading order (\ref{edge-d}).
 \item {\bf Non-universal regime:} $\mu -x\sim O(1/R)$. The spike singularity is resolved according to (\ref{densityPeak1}). This regime is only defined on $S^4$ as the density  depends explicitly on the radius of the sphere, even though $MR$ is still assumed to be large.
\end{itemize}
The first two regimes are universal, in the sense that the results should apply to any sufficiently big contour and do not depend on the $S^4$ compactification (the radius drops out from the expressions for the density). The last regime requires that the theory is compactified  on the sphere, and the computations in that case only apply to the Wilson loop running along the big circle of $S^4$.

\section{Antisymmetric representations}

The Wilson loop in the antisymmetric representation is expressed through the Fermi distribution, for which $1/L$ plays the r\^ole of temperature. When $L$ is large, the effective temperature is low. Then (\ref{cp<->dens}) and (\ref{FreeE}) can be simplified:
\begin{equation}\label{low-T}
 \int_{\nu }^{\mu }dx\,\rho (x)=f
\end{equation}
and 
\begin{equation}\label{int rho x}
 \F^-(\nu )= \int_{\nu }^{\mu }dx\,\rho (x)x.
\end{equation}
This approximation is justified when the scale of variation of the density is much larger than $1/L$, which is true in the bulk and endpoint regimes, because $\mu L\gg1$ and $ML\gg 1$, but not in the non-universal regime when the density varies on scales of order $1/R=2\pi /L$. 

\subsection{Bulk regime}

We begin with the bulk regime, when the density is the same as in the Gaussian matrix model. 
In that case the answer is the same as in $\mathcal{N}=4$ SYM up to rescaling of $\lambda$ by a factor of $ML/2\pi $, and it reads \cite{Yamaguchi:2006tq,Hartnoll:2006is}
\begin{equation}\label{W-}
 \frac{1}{NML}\,\ln W_k^-=\frac{\sqrt{\lambda }}{3\pi ^2}\,\sin^3\theta 
\end{equation}
where $\theta $, defined as $\cos\theta =\nu /\mu $, is the solution of a transcendental equation
\begin{equation}\label{theta}
 \theta -\frac{1}{2}\,\sin 2\theta =\pi f.
\end{equation}

The correction term to the bulk density in (\ref{bulk-density}) is of order $O(\lambda ^{-1/4})$, but contributes to the Wilson loop vev at relative order $\lambda ^{-3/4}$, because the small-scale variations of the density average to zero.  We are not going to study this contribution in detail, since it is subleading to the effect produced by the so far unknown order $\lambda ^{-1/2}$ correction to the density.

From (\ref{theta}) we see that the bulk regime corresponds to $f=O(1)$. When $f$ becomes small\footnote{or very close to one -- all the formulas are invariant under $f\rightarrow 1-f$ charge conjugation symmetry of antisymmetric representations.}, the Fermi level $\nu $ approaches the edge of the eigenvalue distribution and eventually the endpoint regime sets in.

\subsection{Endpoint regime}

Let us define dimensionless variables $u \equiv (\mu-x)/M$ and $v \equiv (\mu-\nu)/M$. 
It is convenient to express the endpoint distribution \eqref{edge-d} for different intervals of $u$, i.e.
\begin{equation}\label{explicit-endpoint-density}
 \rho(u)=
 \begin{cases} 
  \frac{1}{\pi }\sqrt{\frac{2M}{\mu ^3}}\frac{1}{\sqrt{u}} &\mbox{if } u \in (0,1]\\
  \frac{1}{\pi }\sqrt{\frac{2M}{\mu ^3}} \left(\frac{1}{\sqrt{u}}+\frac{1}{\sqrt{u-1}}\right) &\mbox{if } u \in (1,2]\\
  \quad \vdots\\
  \frac{1}{\pi }\sqrt{\frac{2M}{\mu ^3}} \sum _{k=0}^n \frac{1}{\sqrt{k+u-n}} &\mbox{if } u \in (n , n+1]
 \end{cases}
\end{equation}
It is then clear that $f$ should scale as $(M/\mu )^{3/2}\sim \lambda ^{-3/4}$. To make this scaling manifest we introduce
\begin{equation}\label{scaling-of-f-endpoint}
 f=\frac{1}{\pi }\left(\frac{2M}{\mu }\right)^{\frac{3}{2}}\tilde{f}
 =4\sqrt{\pi }\,\lambda ^{-\frac{3}{4}}\tilde{f},
\end{equation}
and assume that $\tilde{f}\sim O(1)$.

\begin{figure}[t]
\begin{center} 
 \centerline{\includegraphics[width=10 cm]{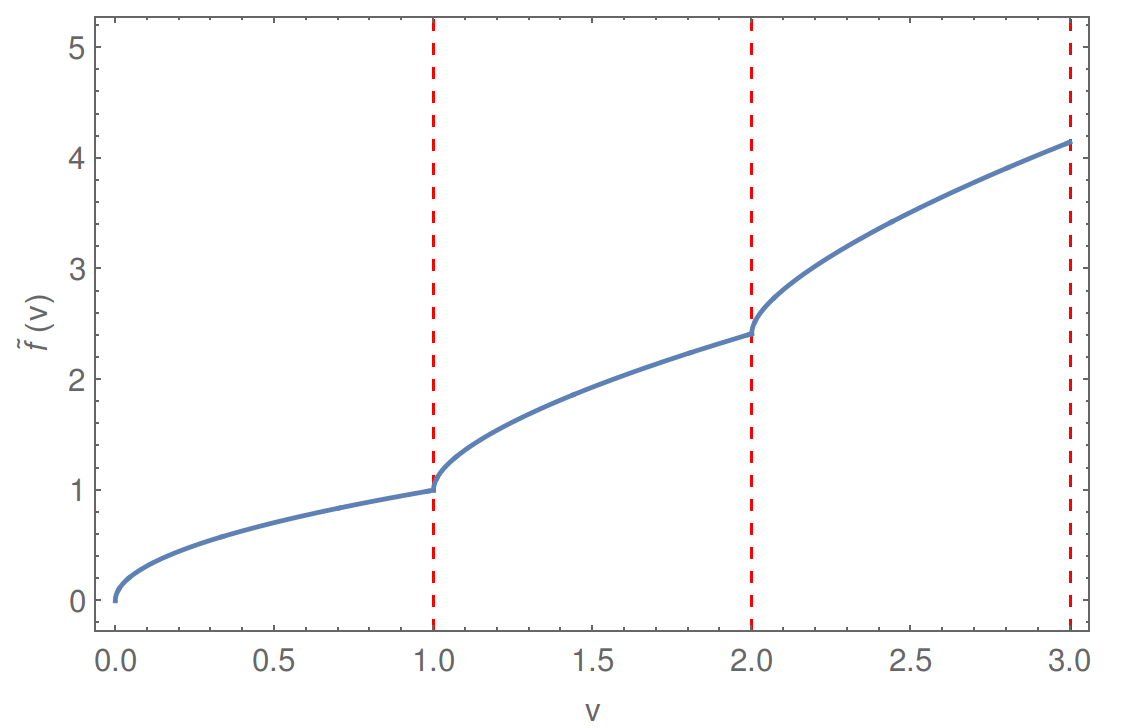}}
\caption{\label{fPlotAntisym}  \small The rescaled function $\tilde{f}(v)$ plotted. 
The dash lines indicate the critical values $\tilde{f}_{\text{critical},n}$ for $n=1,2,3$.  }
\end{center}
\end{figure}

The integral in \eqref{low-T} can be easily performed by summing over the contribution of different intervals of $u$ that are smaller than $v$. For example,
for  $n-1< v \leq n $, with integer $n$, we have:
\begin{equation}
 \frac{1}{M}\,f_n(v)=\int_{0}^{1} \rho (u) \, du + \cdots +
 \int_{n-1 }^{v} \rho (u) \, du
\end{equation}
where we have labeled $f$ with a subindex $n$ as a reminder of the domain of $v$.
For the rescaled variables, the explicit forms are:
\begin{eqnarray} 
 \tilde{f}_1(v)&=&\sqrt{v } \\ 
 \tilde{f}_2(v)&=&\sqrt{v }+\sqrt{v -1} \\ 
 \tilde{f}_3(v)&=& \sqrt{v}+\sqrt{v -2 }+\sqrt{v -3}, 
\end{eqnarray}
and so on. The function $\tilde{f}(v)$ is continuous but discontinuous in its first derivative at the critical points, which are its values at integer points, i.e. $\tilde{f}_{\text{critical},n}~=~\tilde{f}_n(n)$:
\begin{equation}\label{critical-points}
  \tilde{f}_{\text{critical},n} = \left\{1, 1+\sqrt{2}, 1+\sqrt{2}+\sqrt{3}, \ldots \right\}.
\end{equation}
Fig.~\ref{fPlotAntisym} shows a plot of this function for $0<v\leq3$.

The free energy at the saddle point is given by \eqref{int rho x}. 
We are interested in expressing it in terms of the free parameter $\tilde{f}$, therefore, first, we must invert the function $\tilde{f}(v)$. 
The inverse functions of $\tilde{f}_1(v)$ and $\tilde{f}_2(v)$ admits a simple  closed form: 
\begin{eqnarray}
 0<v \leq 1: \quad v_1(\tilde{f})&=&\tilde{f}^2 \label{v1}\\
 1<v \leq 2: \quad v_2(\tilde{f})&=&\frac{\left(\tilde{f}^2+1\right)^2 }{4\tilde{f}^2} \label{v2}
\end{eqnarray}
$v_3(\tilde{f})$ can be also expressed in radicals, but the expression is lengthy, and we do not display it here. 

Now, in terms of $v(\tilde{f})$, and using the saddle point equation \eqref{low-T} reexpressed in the endpoint variables $u$ and $v$, 
the free energy as a function of $f$ is:
\begin{equation}
 \F^-(f) = \mu f -M^2\int _0^{v(\tilde{f})}\; du \; \rho(u) \; u
 \end{equation}
At strong coupling the first term is dominant, and to the first approximation the free energy is simply $\mathcal{F}^-=\mu f=\mu k/N$. The Wilson loop in the rank-$k$ representation therefore behaves as $\,{\rm e}\,^{\mu Lk}$, which is just the $k$-th power of the fundamental representation, as can be expected for small representations on account of the large-$N$ factorization. The second term is of relative order $O(M/\mu )=O(1/\sqrt{\lambda })$ and in the dual holographic description can be interpreted as a quantum correction on the D-brane worldvolume. This term has singularities at the critical points (\ref{critical-points}), which means that the worldvolume effective field theory may have a non-trivial phase structure even at infinitely strong coupling.

\begin{figure}[t]
\begin{center}
 \centerline{\includegraphics[width=10 cm]{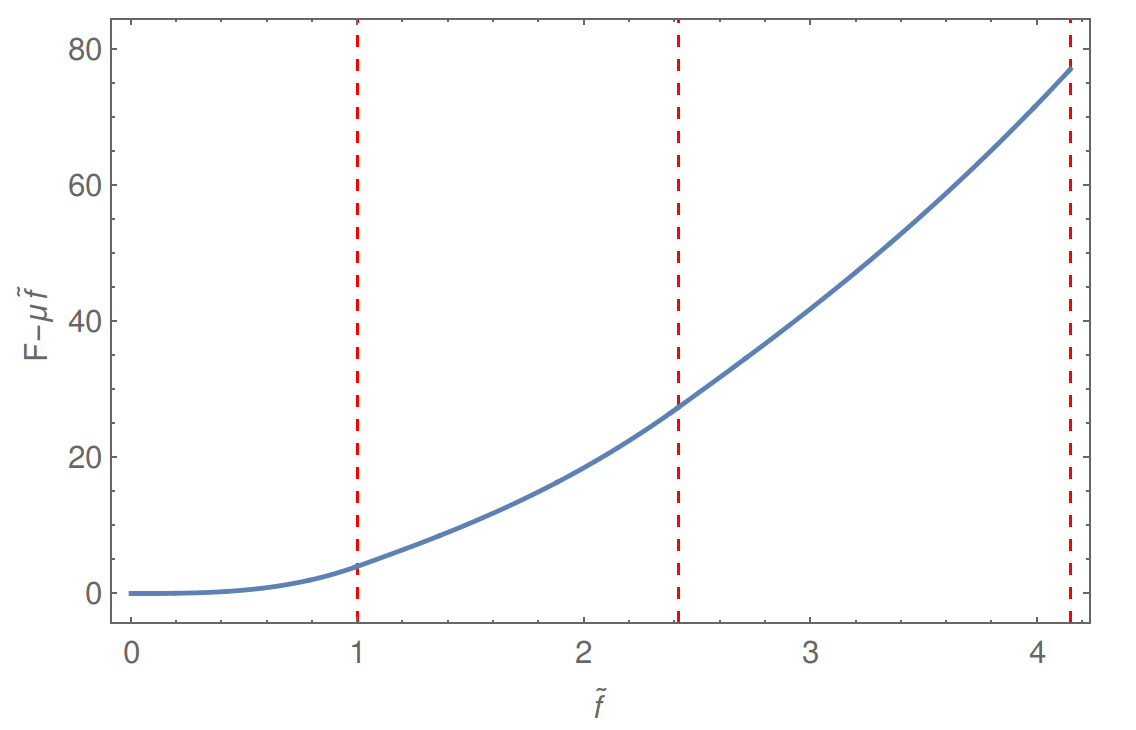}}
\caption{\label{FplotAntisym} \small The subleading term of the free energy plotted, rescaled according to \eqref{free--}.
The dash lines indicate the critical values $\tilde{f}_{\text{critical},n}$ for $n=1,2,3$. }
\end{center}
\end{figure}

Therefore, the phase structures are smooth, due to the subleading term, see fig.~\ref{FplotAntisym}. 

The phase transitions are of the second order, as can be seen by examining the derivatives of free energy. Using that $df/dv=M\rho (v)$, we find:
\[
\frac{d\mathcal{F}^-}{df}=\mu -Mv,\qquad \frac{d^2\mathcal{F}^-}{df^2}
=-\frac{1}{\rho (v)}\,.
\]
The inverse density is singular for integer values of $v$, as can be seen in fig.~\ref{fig-density-edge}, and takes a finite value to the left of the critical point going to zero to the right. The second derivative of the free energy consequently experiences a finite jump across the phase transition. We can see this explicitly by computing the free energy on the first two intervals:
\begin{equation}\label{free--}
 \mathcal{F}^--\mu f=-\frac{\pi M}{3  \lambda^{3/4}}\times 
\begin{cases}
 4\tilde{f}^3 & {\rm }0<\tilde{f}\leq 1
\\
 \tilde{f}^3+6\tilde{f}-\frac{3}{\tilde{f}} & {\rm } 1<\tilde{f}\leq 1+\sqrt{2}
 \\
 \ldots &
\end{cases}
\end{equation}
And it is straightforward to check that $\mathcal{F}^-$ is continuous at $\tilde{f}=1$ together with its first derivative, while the second derivative experiences a finite jump.

\subsection{Non-universal regime}\label{non-universal:sec}

Extremely close to the endpoints, the first peak is resolved,  and we need to solve \eqref{cp<->dens} with the density from \eqref{densityPeak1}. 
From \eqref{cp<->dens} we can see that the Wilson loop enters this regime at 
\[
 f\sim \dfrac{M }{R^{1/2} \mu^{3/2}}\sim (M R)^{-1/2}\lambda^{-3/4},
\]
which is an additional factor $\sqrt{MR}$ smaller compared to (\ref{scaling-of-f-endpoint}).

Considering again symmetric and antisymmetric representations in parallel, we can see that the free energy scales as
\begin{equation}
 \mathcal{F}^\pm=\mu f+\dfrac{\sqrt{2 \pi} M}{\pi^{4}(M R)^{3/2} \lambda^{3/4}}\,h^\pm\left( \pi^3 \sqrt{\frac{M R}{8 \pi}}\lambda^{3/4} f\right).
 \label{freeEnergy}
\end{equation}
The scaling function $h^\pm(\f)$ can be written in the parametric form:
\begin{eqnarray}
 h^\pm(\tt{f})&=&-\f\ln q\mp\int_{0}^{\infty }\frac{d\kappa }{\kappa ^{\frac{3}{2}}}\,\,\cos2\pi \kappa \sin^2\pi \kappa \,\Gamma ^2(\kappa )
 \,{\rm e}\,^{-2\kappa \left(\ln\kappa -1\right)}
\nonumber \\
&&\times 
 \left[\ln\left(1\mp q\right)\pm\frac{q}{1+\kappa }\,
 {}_2F_1(1,1+\kappa ;2+\kappa ;\pm q)\right]
 \\
 \label{f--f}
 \f&=&q\int_{0}^{\infty }\frac{d\kappa }{\sqrt{\kappa }}\,\,\cos2\pi \kappa \sin^2\pi \kappa \,\Gamma ^2(\kappa )
 \,{\rm e}\,^{-2\kappa \left(\ln\kappa -1\right)}
\nonumber \\
&&\times \frac{1}{1+\kappa }\,
 {}_2F_1(1,1+\kappa ;2+\kappa ;\pm q).\label{slash-f}
\end{eqnarray}
The parameter $q$ is the fugacity variable related to the chemical potential $\nu $ in (\ref{cp<->dens}), (\ref{FreeE}) by $q=\,{\rm e}\,^{2\pi R(\mu -\nu )}$. For the antisymmetric representations (fermions), $q$ changes from zero (low density) at to infinity (high density), while for bosons the chemical potential must be negative and so $q$ is always smaller than one. 

The low density approximation, $q\rightarrow 0$, corresponds to Boltzmann statistics, when the difference between symmetric and antisymmetric representation disappears. The hypergeometric function in the above equations can then be replaced by $1$. We thus get:
\begin{equation}
 \f=\pi ^3\,{\rm e}\,^{-2}q\qquad \left(q\rightarrow 0\right),
\end{equation}
and 
\begin{equation}
 h^\pm(\f)=\f\left(1-\ln q\right)=\f\ln\frac{\pi ^3}{\,{\rm e}\,^{}\f}\qquad \left(\f\rightarrow 0\right).
\end{equation}
For the Wilson loop we get the following result, for both symmetric and antisymmetric representations:
\begin{equation}
 W_k^\pm\simeq \left[\frac{N}{k}\,\sqrt{\frac{8\pi }{MR}}\,
 \lambda ^{-\frac{3}{4}}\,{\rm e}\,^{\left(\sqrt{\lambda }-\pi \right)RM-1}\right]^k\simeq \frac{W_1^k}{k!}\,,
\end{equation}
where $W_1$ is the expectation of the Wilson loop in the fundamental representation that at strong coupling behaves as \cite{Chen:2014vka}:
\begin{equation}
 W_1=N\sqrt{\frac{8\pi }{MR}}\,
 \lambda ^{-\frac{3}{4}}\,{\rm e}\,^{\left(\sqrt{\lambda }-\pi \right)RM-2}.
\end{equation}
This is indeed the correct behavior that we expect at very small $k\ll N$. The Wilson loop  then picks the leading contribution from the term with the largest number of traces, due to the large-$N$ factorization.
The result above follows from
\begin{equation}\label{large-trace}
 \mathop{\mathrm{tr}}\nolimits_{\mathcal{R}_k^\pm}U=\frac{1}{k!}\,\left(\mathop{\mathrm{tr}}U\right)^k+O\left(\mathop{\mathrm{tr}}\nolimits^{k-1}\right).
\end{equation}

In the high-density regime, symmetric and antisymmetric representations behave very differently. For fermions, $\f$ grows indefinitely with $q$ and asymptotes to $\f\simeq 2\pi ^{5/2}\sqrt{\ln q}$ at large $q$. The function $h^-(\f)$ becomes negative and its absolute value grows as the cube of the argument:
\begin{equation}
 h^-(\f)\simeq -\frac{\f^3}{12\pi ^5}\qquad \left(\f\rightarrow \infty \right),
\end{equation}
which can be seen to match (\ref{free--}) in the overlap of the region between the two regimes. 

In the bosonic case, $q$ cannot be bigger than one. When $q$ approaches one, the integral  in (\ref{f--f}) converges to a finite value, indicating that the solution for $q$ exists only for sufficiently small $\f$. This corresponds to the  Bose-Einstein condensation in the analog statistical system. 
 We discuss how to compute the symmetric-representation Wilson loop for arbitrary $f$  in the next section.

\section{Symmetric representations}\label{symmrep-sec}

\subsection{Bose-Einstein condensation}

The Bose-Einstein condensation for symmetric-representation Wilson loops has been discussed for the Gaussian model  \cite{Hartnoll:2006is} of $\mathcal{N}=4$ SYM \cite{Erickson:2000af,Drukker:2000rr},
 as well as in the unitary matrix models \cite{Grignani:2009ua}. 
To continue the Wilson loop expectation values past the transition point, two prescriptions have been used: 
(i) analytic continuation in $f$ \cite{Hartnoll:2006is,Grignani:2009ua} 
and (ii) observation that for sufficiently large $k$ the rank-$k$ symmetric representations become equivalent to the $k$-wrapped fundamental representation \cite{Yamaguchi:2007ps}:
\begin{equation}
 \left\langle \mathop{\mathrm{tr}}\nolimits_{\mathcal{R}^+_k}\,{\rm e}\,^{L\Phi _0}\right\rangle\simeq \left\langle \mathop{\mathrm{tr}}\,{\rm e}\,^{kL\Phi _0}\right\rangle\qquad \left({\rm for~sufficiently~large~}k\right).
\end{equation}
The answer for multiply-wrapped Wilson loops is known exactly at any $N$ and $k$,  in the Gaussian model  \cite{Drukker:2000rr}. The scaling limit $N\rightarrow \infty $, $k\rightarrow \infty $ can be obtained as an approximation to this exact answer \cite{Drukker:2005kx}, 
and it agrees with the symmetric-representation Wilson loop
analytically continued far past the Bose-Einstein condensation point. The equivalence of the two approaches, however, has never been actually proven, and since our model is not Gaussian, we prefer to derive all the necessary results from scratch.

Technically, the condensation happens because the hypergeometric function in (\ref{slash-f}) has a logarithmic branch point at $q=1$:
\begin{equation}
 \frac{1}{1+\kappa }\,\label{hyperlog}
 {}_2F_1(1,1+\kappa ;2+\kappa ;\pm q)= -\ln(1-q)+\gamma +\psi (1+\kappa)+O(1-q).
\end{equation}
However, this entails no divergences in $\f$, because
the
 logarithmic term integrates to zero in (\ref{slash-f}) and $\f$  approaches a finite limiting value
\begin{equation}
 \f_c\equiv \left.\f\right|_{q=1}=\pi ^3\sum_{n=1}^{\infty }\frac{n^{2n-\frac{1}{2}}\,{\rm e}\,^{-2n}}{n!^2}=11.88\ldots 
\end{equation}
If $\f$ exceeds $\f_c$,  solution to the saddle-point equation ceases to exist.

\begin{figure}[t]
\begin{center}
 \centerline{\includegraphics[width=8cm]{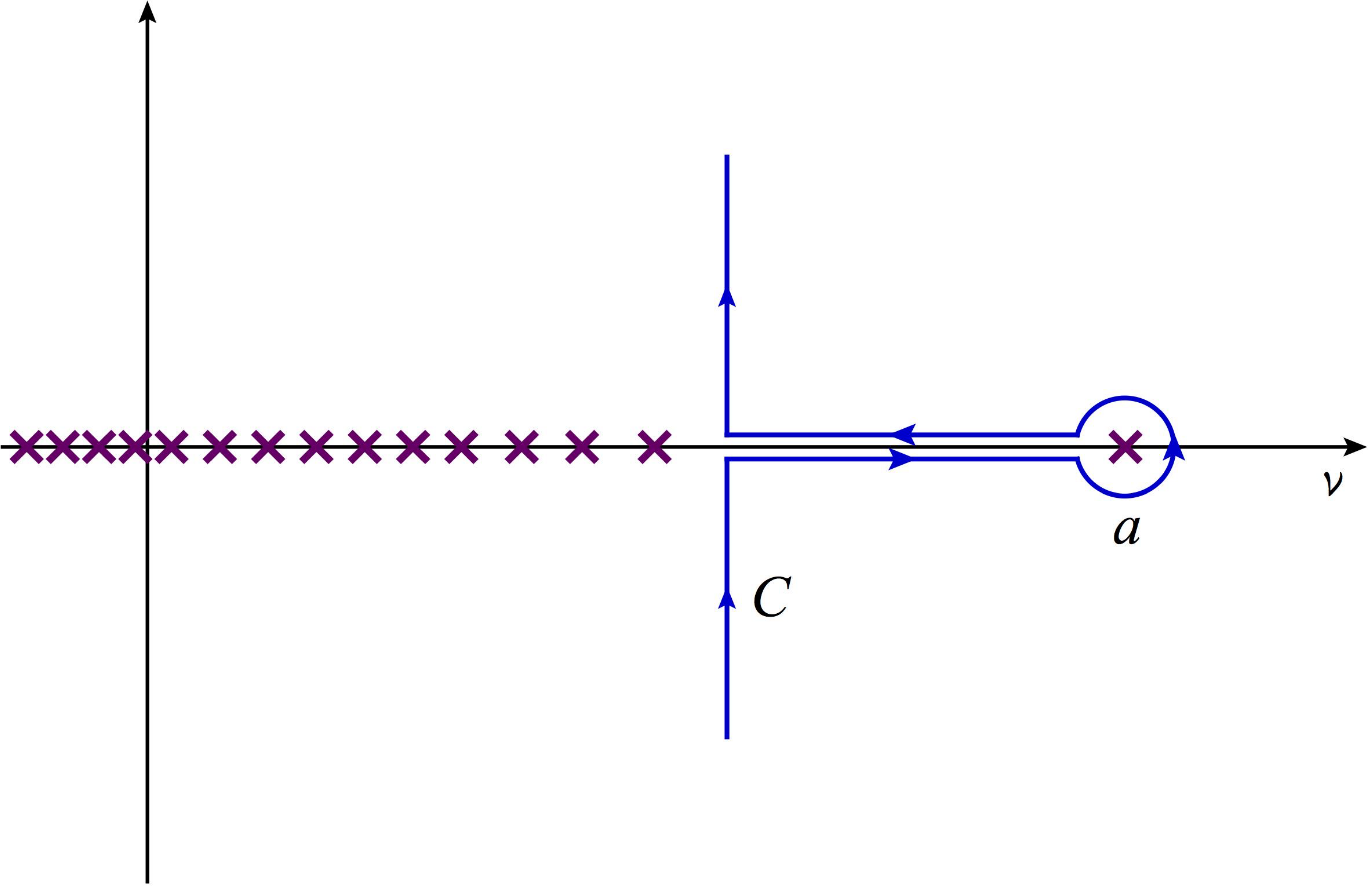}}
\caption{\label{Contour:fig}\small The contour deformation in (\ref{a-average}).}
\end{center}
\end{figure}

This happens because the chemical potential of a non-interacting Bose gas cannot be positive, 
which translates into $\nu$ being bigger than $\mu $ in our case. 
The edge behavior of the density of states, $\rho (x)\sim \sqrt{\mu -x}$, guarantees the convergence of the integral in (\ref{cp<->dens}) at $\nu =\mu $. 
Following the analogy with statistical mechanics, we expect that the excess density $\f-\f_c$ will condense in the ground state, 
identified in our case with the largest eigenvalue $\max\left\{a_j\right\}\equiv a $. 
An important difference to the textbook treatment is the randomness of the energy levels which themselves are integration variables with the measure defined by the eigenvalue integral (\ref{Zed}).

To take into account the condensate,  we need to deform the contour of integration in (\ref{a-average}) as shown in fig.~\ref{Contour:fig}, picking the pole at $\nu =a $:
\begin{equation}
 W_k^+=\left\langle \,{\rm e}\,^{kLa }\prod_{a_j\neq a }^{}\frac{1}{1-\,{\rm e}\,^{L\left(a_j-a \right)}}\right\rangle+\ldots 
\end{equation}
The remaining contour integral is exponentially small compared to the pole contribution, and can be omitted at large $N$. The average over the bulk eigenvalues can then be replaced by convolution with the saddle-point density, while the integral over the largest eigenvalue should be kept as it is:
\begin{equation}\label{one-eigenvalue-int}
 W^+_k \simeq  \int_{\mu }^{\infty } da \, P(a)\exp\left[kLa-N\int_{-\mu }^{\mu }dx\,\rho (x)\ln\left(1-\,{\rm e}\,^{L(x-a)}\right)\right].
\end{equation}
Here $P(a)$ is the probability to find the largest eigenvalue at point $a$ outside the eigenvalue interval:
\begin{equation}
 P(a)=\,{\rm e}\,^{-\frac{2NL^2}{\lambda }\,a^2+NL\mathcal{F}_0}\prod_{j}^{}\mathcal{Z}_{\rm 1-loop}(a-a_j),
\end{equation}
where
\begin{equation}
 \mathcal{F}_0=\frac{2L\mu ^2}{\lambda }-\frac{1}{L}\int_{-\mu }^{\mu }dx\,\rho (x)\ln\mathcal{Z}_{\rm 1-loop}(\mu -x).
\end{equation}
The constant $\mathcal{F}_0$ is the free energy of the eigenvalue sitting right at the edge of the bulk distribution. 
The probability to find an eigenvalue at $a$ is determined by the energy cost of taking an eigenvalue from the edge and moving it to $a$, and it is given by the difference in free energies.

The probability to pull out one eigenvalue outside the bulk distribution is exponentially small, 
but in the integral (\ref{one-eigenvalue-int}) the negative exponent  is counteracted by the positive exponent in the thermodynamic weight. 
The largest contribution to the integral hence comes from the saddle point:
\begin{equation}
 \ln W_k^+=NL\mathcal{F}(a)+O(N^0).
\end{equation}
This is similar to (\ref{FreeE}), but now the free energy  is given by
 \begin{equation}\label{FreeESym}
  \F(a) = \mathcal{F}_0+f a+\frac{1}{L}\int _{-\mu}^\mu dx \rho (x) \ln\frac{\mathcal{Z}_{\rm 1-loop}(a-x)}{1-\,{\rm e}\,^{L(x-a)}}-\frac{2L}{\lambda }\,a^2.
  \end{equation}
 The position of the saddle point is determined by the following equation:
 \begin{equation}\label{f}
f = \dfrac{4L}{\lambda}\, a - \frac{2}{L} \int_{-\mu }^{\mu } dx\,\rho (x)\left[ S(a-x)-\frac{1}{\,{\rm e}\,^{L(a-x)}- 1}\right]
 \end{equation}
with the kernel:
\begin{equation}
S(x)\equiv \dfrac{1}{x} + \dfrac{1}{2} \,K(x+M)+\frac{1}{2}\,K(x-M)-K(x),
\end{equation}
where we used the explicit form on the one-loop measure (\ref{Z1loop}), and defined
\begin{equation}
 K(x)=-\frac{H'(x)}{H(x)}\,.
\end{equation}
The first two terms in (\ref{f}) were absent in (\ref{cp<->dens}) and can be interpreted as the density of the condensate.

Another way to calculate the Wilson loop for $\f>\f_c$ is to continue the free energy analytically from $\f<\f_c$. Indeed, the logarithmic singularity of (\ref{hyperlog}) cancels upon integration in (\ref{slash-f}) and $\f(q)$, as a function of $q$, has no singularities at $q=1$. We prove in appendix~\ref{a-cont:sec} that the result of the analytic continuation gives the same results as the calculation above under fairly general conditions, applicable in particular to the model under consideration. The Bose-Einstein condensation consequently does not lead to any singularities in the expectation value of the Wilson loop. 

In the universal regime $L(a-\mu )\gg 1$, the thermal contribution to the free energy (the last term in (\ref{f})) becomes exponentially suppressed and can be dropped:
\begin{eqnarray}
 \mathcal{F}(a)&\simeq& \mathcal{F}_0+f a+\frac{1}{L}\int _{-\mu}^\mu dx \rho (x) \ln\mathcal{Z}_{\rm 1-loop}(a-x)-\frac{2L}{\lambda }\,a^2
 \\
f &\simeq & \dfrac{4L}{\lambda}\, a - \frac{2}{L} \int_{-\mu }^{\mu } dx\,\rho (x) S(a-x).
\label{integral-for-f}
\end{eqnarray}
In this approximation, the  $k$-symmetric and $k$-wrapped Wilson loops have the same expectation value:
\begin{equation}
 W_k^+\sim \left\langle \mathop{\mathrm{tr}}\,{\rm e}\,^{kL\Phi _0}\right\rangle,
\end{equation}
because the right-hand side is calculated by equation (\ref{one-eigenvalue-int}) without the second term in the exponent \cite{Yamaguchi:2006te,Yamaguchi:2007ps}. 
This result contradicts intuition based on the large-$N$ factorization, as the Wilson loop now picks the largest contribution from the term with the smallest number of traces.

\subsection{Bulk regime}

When $a-\mu \sim \mu $, the density is the Wigner distribution and the kernel $S(x)$ can be approximated by $M^2R^2/x$ \cite{Buchel:2013id}. The answer is then obtained 
from the Wilson loop in the Gaussian model \cite{Drukker:2005kx,Hartnoll:2006is} by simple rescaling: 
\begin{equation}\label{W+}
 \ln W_k^+ = \frac{NL^2M^2}{2\pi ^2}\left(\kappa\sqrt{1+\kappa^2}
 +\mathop{\mathrm{arcsinh}}\kappa 
 \right),
\end{equation}
where
\begin{equation}
 \kappa \equiv \frac{\lambda f}{4L\mu }=\frac{\pi \sqrt{\lambda }\,f}{2ML}\,.
\end{equation}
It is interesting to notice that  the consistency of this regime requires the rank of representation to scale linearly with the size of the contour. There should be a simple explanation to this fact in the dual supergravity picture.

\subsection{Endpoint regime}

We now consider the case when the saddle point $a$ approaches the endpoint of the eigenvalue distribution. As before, we introduce the dimensionless variables
\begin{equation}
 v=\frac{a-\mu }{M}\,,\qquad u=\frac{\mu -x}{M}\,.
\end{equation}
The density near the endpoint has the form (\ref{explicit-endpoint-density}). It cannot be just substituted into (\ref{integral-for-f}) because the integral would then diverge. The strategy is to separate the bulk and endpoint contributions to $f$ as follow:
\begin{equation}
 f\equiv f_{\infty}+f_{\text{ep}},
\end{equation}
where $f_{\infty }$ is computed with the Wigner density and with $S(x)$ approximated by $M^2R^2/x$:
\begin{equation}
f_{\infty} = \dfrac{4 L}{\lambda} a - \dfrac{M^2L}{2\pi^2 } \int_{-\mu }^{\mu } dx\,\rho_\infty (x) \dfrac{1}{a-x}=\frac{4L}{\lambda }\,\sqrt{a^2-\mu ^2}\approx 
 \frac{4LM}{\sqrt{\pi }\,\lambda ^{\frac{3}{4}}}\,\sqrt{v}.
 \end{equation}
Notice that the rank of representation scales linearly with the length of the contour as in the bulk regime, but its coupling dependence is different: $f\sim \lambda ^{-3/4}$ instead of $\lambda ^{-1/2}$. We thus introduce the rescaled variable
\begin{equation}
 \tilde{f}=\frac{\sqrt{\pi }\lambda ^{\frac{3}{4}}}{4LM}\,f,
\end{equation}
such that
\begin{equation}\label{tildefinfinity}
 \tilde{f}_\infty =\sqrt{v}.
\end{equation}

For the genuine endpoint contribution we find:
\begin{equation}
 f_{\rm ep}=\dfrac{2}{L } \int_{-\mu }^{\mu } dx\,
 \left(
 \rho_\infty (x) \dfrac{M^2R^2}{a-x}- \rho (x) S(a-x).
 \right).
\end{equation}
In spite of proximity to the endpoint, the argument of the kernel is still very big: $a-x=M(u+v)\gg 1$, which justifies the use of an approximate expression
\begin{equation}
 S(a-x)\approx MR^2\tilde{S}(u+v)
\end{equation}
with
\begin{equation}
 \tilde{S}(w)=\frac{1}{2}\,\left(w+1\right)\ln(w+1)^2+\frac{1}{2}\,\left(w-1\right)\ln(w-1)^2-w\ln w^2.
\end{equation}
Writing the endpoint density (\ref{edge-d}), (\ref{explicit-endpoint-density}) as
\begin{equation}
 \rho (u)=\frac{1}{\pi }\sqrt{\frac{2M}{\mu ^3}}\,F(u),
\end{equation}
and introducing the function
\begin{equation}\label{Xv}
 X(v)=\int_{0}^{\infty }\frac{du}{2\pi }\left(\frac{2\sqrt{u}}{u+v}-\tilde{S}(u+v)F(u)\right),
\end{equation}
we find that
\begin{equation}
 \tilde{f}_{\rm ep}=X(v).
\end{equation}

The function $X(v)$ can be computed by noticing that
\begin{equation}
 \tilde{S}''(w)=\frac{1}{w+1}+\frac{1}{w-1}-\frac{2}{w}\,,
\end{equation}
and differentiating (\ref{Xv}) twice, after which elementary integrations yield:
\begin{equation}
 X''(v)=\frac{1}{2\sqrt{v}}-\frac{\theta (v-1)}{2\sqrt{v-1}}+\frac{1}{4}\,v^{-\frac{3}{2}}\,.
\end{equation}
We thus get:
\begin{equation}\label{func-X}
 X(v)=\frac{2}{3}\,v^{\frac{3}{2}}-\frac{2}{3}\,\theta (v-1)\left(v-1\right)^{\frac{3}{2}}-\sqrt{v}\,,
\end{equation}
where the integration constants are set to zero in order to get the correct behavior at $v\rightarrow \infty $. Interestingly, the function $X(v)$ has a branch-point singularity outside the eigenvalue support, again due to the resonance appearing in the kernel $S(x)$.
Combining this result with (\ref{tildefinfinity}), we find:
\begin{equation}
 \frac{3}{2}\,\tilde{f}=v^{\frac{3}{2}}-\theta (v-1)\left(v-1\right)^{\frac{3}{2}}.
\end{equation}
An alternative derivation based on the Wiener-Hopf method  \cite{Chen:2014vka}  is given in the appendix~\ref{computeX:sec}.

Recalling that the equation for $f$ was obtained by differentiating the free energy with respect to $a$, we can get the free energy by integration:
\begin{eqnarray}
 \mathcal{F}-\mu f&=&\frac{4LM^2}{\sqrt{\pi }\,\lambda ^{\frac{3}{4}}}
 \left(\tilde{f }v-\int_{0}^{v}dw\,\tilde{f}(w)\right)
\nonumber \\
&=&
 \frac{8LM^2}{5\sqrt{\pi }\,\lambda ^{\frac{3}{4}}}\left[
 v^{\frac{5}{2}}+\left(v+\frac{2}{3}\right)\theta (v-1)\left(v-1\right)^{\frac{3}{2}}
 \right].
\end{eqnarray}
As before, we would like to express the free energy in terms of $\tilde{f}$. We need then to obtain the inverse function $v(\tilde{f})$. For $v<1$ or equivalently $\tilde{f}<2/3$, the answer is very simple:
\begin{equation}
 v=\left(\frac{3\tilde{f}}{2}\right)^{\frac{2}{3}},\qquad 
 \mathcal{F}-\mu f= \frac{12^{\frac{5}{3}}LM^2}{5\sqrt{\pi }\,\lambda ^{\frac{3}{4}}}\,\tilde{f}^{\frac{5}{3}}\qquad \left(\tilde{f}<\frac{2}{3}\right).
\end{equation}
At the critical point 
$$\tilde{f}_c=\frac{2}{3}$$ 
the free energy experiences a phase transition. It is easy to show that $\mathcal{F}$ is continuous together with its two first derivatives, while the third derivative diverges at the critical point:
\begin{equation}
 \mathcal{F}={\rm regular}+\,{\rm const}\,\left(\frac{2}{3}-\tilde{f}\right)^{\frac{5}{2}}.
\end{equation}
The phase transition is consequently of the third order.

\subsection{Non-universal regime}

The general scaling dependence of the variable $f$ and of the free energy on $MR$ and $\lambda $ have already been discussed in sec.~\ref{non-universal:sec}:
\begin{eqnarray}
 f&\sim& (M R)^{-1/2} \lambda^{-3/4} 
\nonumber \\
\mathcal{F}-\mu f&\sim& M\left(MR\right)^{-3/2}\lambda ^{-3/4}.
\end{eqnarray}
Explicit integral representation for the free energy can be obtained by plugging the scaling form of the density (\ref{densityPeak1}) into (\ref{FreeESym}), (\ref{f}). The Fourier representation of the the resulting integrals is given in appendix~\ref{computeX:sec}.

\section{Conclusions}

The leading-order strong-coupling solution of the SYM* matrix model is given by the same Wigner distribution as in the $\mathcal{N}=4$ SYM, 
up to rescaling of the coupling $\lambda \rightarrow \lambda L^2/4\pi ^2$. 
Wilson loop expectation values in the bulk regime, as a consequence, are obtained from the Gaussian results  by simple rescaling. 
This is quite surprising, as the dual geometry of SYM* is rather different from $AdS_5\times S^5$ away from the boundary.  
It would be very interesting to verify these basic matrix-model predictions  by finding classical D-branes solutions in the Pilch-Warner geometry.

Even more interesting phenomena occur for parametrically small representations. The leading order then  is just 
\begin{equation}
 W^\pm_k\sim \,{\rm e}\,^{\sqrt{\lambda }\,kML},
\end{equation}
the result expected form the large-$N$ factorization. It is the next, $\sqrt{\lambda }$ suppressed term that displays a non-trivial behavior, 
with phase transitions in the representation parameter $f=k/N$, appropriately rescaled. In the dual picture, $f$  maps to the density of electro-magnetic flux on the D-brane worldvolume. Perhaps the non-analytic behavior of the Wilson loops  can be interpreted as phase transitions in the effective field theory on the D-brane worldvolume with the electric or magnetic flux playing the r\^{o}le of tunable parameter. These phase transitions and quantum phase transitions in the matrix  model at finite but large $\lambda $ have the same origin.

We have mainly concentrated on the infinite-coupling regime of the theory. The discontinuities in the higher-rank Wilson loops should persist at finite coupling as soon as there are resonance peaks in the density\footnote{We would like to thank J.~Russo for this remark.}, which is the case for $\lambda >\lambda _c^{(1)}\approx 35.4$ \cite{Russo:2013qaa}. It would be interesting to completely map the phase diagram of Wilson loop expectation values in the $\lambda -f$ plane.

\subsection*{Acknowledgements}

We would like to thank J.~Minahan for discussions and J.~Russo for comments on the manuscript.
K.Z. gratefully acknowledges support from the Simons Center for Geometry and Physics, Stony Brook University at which some of the research for this paper was performed.
This work was supported by the Marie
Curie network GATIS of the European Union's FP7 Programme under REA Grant
Agreement No 317089, by the ERC advanced grant No 341222
and by the Swedish Research Council (VR) grant
2013-4329. 

\appendix

\section{Analytic continuation}\label{a-cont:sec}

Consider an eigenvalue model defined by the partition function
\begin{equation}
 Z=\int_{}^{}d^{N-1}a\,\prod_{i<j}^{}\mathcal{Z}(a_i-a_j)\,{\rm e}\,^{-N\sum_{i}^{}V(a_i)}.
\end{equation}
The eigenvalue density for this model satisfies the saddle-point equation
\begin{equation}\label{mamosadle}
 \strokedint_{-\mu }^{\mu }dy\,R(x-y)\rho (y)=V'(x)
\end{equation}
with
\begin{equation}
 R(x)=\frac{\mathcal{Z}'(x)}{\mathcal{Z}(x)}\,.
\end{equation}
We assume that 
\begin{equation}\label{poleinR}
 R(x)=\frac{2}{x}+{\rm regular},
\end{equation}
such that the kernel in the saddle-point equation is of the Hilbert type, and the density has the usual square-root asymptotics near the end-points. It will be important for the subsequent derivation that the residue of the kernel is exactly two. If it is different from two, the analytic continuation will not describe Bose-Einstein condensation correctly. 

The $k$-symmetric Wilson loop in this model is determined by the free energy\footnote{Here  we set $L$ to one to simplify the notations. The dependence on $L$ can be easily recovered by dimensional analysis. }
\begin{equation}
 \mathcal{F}(\nu )=f\nu -\int_{-\mu }^{\mu }dx\,\rho (x)\ln\left(1-\,{\rm e}\,^{x-\nu }\right),
\end{equation}
The chemical potential $\nu $ is determined by minimization of the free energy:
\begin{equation}\label{nu->f}
 \mathcal{F}'(\nu )= f-\int_{-\mu }^{\mu }\frac{dx\,\rho (x)}{\,{\rm e}\,^{\nu -x}-1}=0,
\end{equation}
which implicitly determines $\mathcal{F}$ as a function of $f$. These formulas are literally valid for sufficiently small $f<f_c$, such that (\ref{nu->f}) has a real solution. Our goal is to analytically continue the free energy past the critical point.

We first consider the analytic structure of $\mathcal{F}'$ as a function of $\nu $. As such, it has a cut from $-\mu $ to $\mu $ with the discontinuity
\begin{equation}\label{DiscF'}
 \mathop{\mathrm{Disc}}\mathcal{F}'(x)=2i\pi\rho (x). 
\end{equation}
Consequently,  near $\nu =\mu $ the function $\mathcal{F}'(\nu )$ has a double expansion that combines analytic  and square-root terms:
\begin{equation}\label{A8}
 \mathcal{F}'(\nu )=f-f_c+f_1(\nu -\mu )+\ldots +\sqrt{\nu -\mu }\left[g_0+g_1(\nu -\mu )+\ldots \right].
\end{equation}
The free energy is obtained by integration:
\begin{eqnarray}\label{A9}
 \mathcal{F}(\nu )&=&c+(f-f_c)(\nu -\mu )+\frac{1}{2}\,f_{1}(\nu -\mu )^2+\ldots
\nonumber \\
&& 
 +\sqrt{\nu -\mu }\left[\frac{2}{3}\,g_0(\nu -\mu )+\frac{2}{5}\,g_1(\nu -\mu )^2+\ldots \right].
\end{eqnarray}
The constant $g_0$ must be positive, due to positivity of the eigenvalue density.
The solution to (\ref{nu->f}) thus only exists for $f<f_c$. 

When $f_c-f$ is small, (\ref{nu->f}) can be solved iteratively with the help of the expansion (\ref{A8}). Substituting the solution back to $\mathcal{F}(\nu )$, we get to the first approximation:
\begin{equation}
 \mathcal{F}(f)=c-\frac{(f_c-f)^3}{3g_0^2}+\ldots 
\end{equation}
While $\mathcal{F}(\nu )$ has a square-root singularity, the function $\mathcal{F}(f)$ is analytic
and can be continued past $f_c$, which is just an inflection point of the free energy. The analytic continuation in $f$ is equivalent to continuing $\mathcal{F}(\nu )$ in $\nu $ through the square-root branch cut that changes the sign of $f_c-f$ to negative. We denote the result of such analytic continuation by $\tilde{\mathcal{F}}(\nu )$. 

The function $\tilde{\mathcal{F}}(\nu)$ has the same continuous part as $\mathcal{F}(\nu )$ but the opposite discontinuity:
\begin{equation}\label{discontinuityconds}
 \mathop{\mathrm{Cont}}\tilde{\mathcal{F}}(x)=\mathop{\mathrm{\mathcal{F}}}(x),\qquad \mathop{\mathrm{Disc}}\tilde{\mathcal{F}}(x)=-\mathop{\mathrm{Disc}}\mathcal{F}(x),
\end{equation}
and is completely characterized by these two conditions. 

We can construct $\tilde{\mathcal{F}}(\nu )$ by introducing an auxiliary function, the generalized resolvent:
\begin{equation}\label{resolvent}
 W(\nu )=V(\nu )-\int_{-\mu }^{\mu }dy\,\rho (y)\ln\mathcal{Z}(\nu -y)-\mathcal{F}_0,
\end{equation}
where $\mathcal{F}_0$ is just a constant, to be determined later.
The saddle-point equation (\ref{mamosadle}) is equivalent to the two conditions:
\begin{equation}\label{A13}
 \mathop{\mathrm{Cont}}W'(x)=0,\qquad \mathop{\mathrm{Disc}}W'(x)=4i\pi \rho (x) ,
\end{equation}
where the second equation is a consequence of (\ref{poleinR}). This implies, in view of (\ref{DiscF'}), that
\begin{equation}
 \mathop{\mathrm{Cont}}W(x)=0,\qquad \mathop{\mathrm{Disc}}W(x)=2 \mathop{\mathrm{Disc}}\mathcal{F}(x).
\end{equation}
Strictly speaking, the first equation in  (\ref{A13}) only implies that the continuous part of $W(x)$ is a constant independent of $x$, but this constant can be adjusted to zero by setting
\begin{equation}
 \mathcal{F}_0=V(\mu )-\int_{-\mu }^{\mu }dy\,\rho (y)\ln\mathcal{Z}(\mu -y)
\end{equation}
in (\ref{resolvent}). 

It follows from the equations above that
\begin{equation}
 \tilde{\mathcal{F}}(\nu )=\mathcal{F}(\nu )-W(\nu ).
\end{equation}
Indeed, this function satisfies the right discontinuity condition (\ref{discontinuityconds}). Explicitly,
\begin{equation}\label{Free-analytic}
 \tilde{\mathcal{F}}(\nu )=\mathcal{F}_0+f\nu-V(\nu ) +\int_{-\mu }^{\mu }dx\,\rho (x)\ln\frac{\mathcal{Z}(\nu -x)}{1-\,{\rm e}\,^{x-\nu }},
\end{equation}
The function $W(\nu )$ can be interpreted as the energy cost of pulling one eigenvalue out of the bulk of the distribution and placing it at point $\nu $. Comparing (\ref{Free-analytic}) with (\ref{FreeESym}) we see that the analytic continuation of the free energy gives the same result as Bose-Einstein condensation on the largest eigenvalue. The Bose-Einstein condensation thus does not lead to any thermodynamic singularities in the expectation values of the $k$-symmetric Wilson loops.

\section{Wiener-Hopf method}\label{computeX:sec}

In this appendix we derive the general formulas for the symmetric Wilson loop (\ref{FreeESym}), (\ref{f}) from the Wiener-Hopf method developed to solve the saddle-point equations for the matrix model in \cite{Chen:2014vka}, assuming that $a$ is very close to $\mu $.
This covers both endpoint regime and the non-universal regime. As in \cite{Chen:2014vka} we use the units in which $R=1$ in this appendix, and introduce the endpoint variables $\xi =\mu -x$ and $\eta =\mu-a<0$. 

Parametrizing the endpoint distribution as:
\begin{equation}
\rho(x)= \frac{2^{3/2}}{\pi  \mu^{3/2}}F(\xi ),
\end{equation}
we can identify three contributions to $f$ in (\ref{f}):
\begin{equation}
 f\equiv f_{\infty}+f_{\text{ep}}+f_{\text{th}},
\end{equation}
where
\begin{equation}\label{saddle_point_eq_bulk}
f_{\infty} = \dfrac{8\pi }{\lambda} a -  \int_{-\mu }^{\mu } \frac{dx}{\pi }\,\,\rho_\infty (x) \dfrac{1+M^2}{a-x}=\frac{2\left(M^2+1\right)}{ \pi \mu^2 } \sqrt{a^2-\mu^2}\simeq \frac{2^{\frac{3}{2}}\left(M^2+1\right)}{\pi \mu ^{\frac{3}{2}}}\,\sqrt{-\eta }
 \end{equation}
is the bulk contribution,
\begin{equation}
 f_{\text{ep}}=\frac{2^{3/2}}{\pi ^2  \mu^{3/2}} X(\eta ) 
 \end{equation}
 is the end-point contribution and
 \begin{equation}
 f_{\text{th}}=\frac{2^{3/2}}{\pi   \mu^{3/2}} \int _0^{\infty }\frac{F(\xi ) d\xi }{e^{\xi -\eta }-1}
\end{equation}
is the thermodynamic term. Here we concentrate on the endpoint part, defined in terms of the function
 \begin{equation}
  X(\eta ) \equiv \int _0^{\infty }d\xi \, \left[\frac{\left(M^2+1\right) \sqrt{\xi }}{\xi -\eta }-F(\xi ) S(\xi -\eta )\right].
 \end{equation}

Our main observation  is that $X(\eta )$ is the same function that was introduced in \cite{Chen:2014vka} to solve the integral equation of the matrix model, where it plays the role of the remainder function in the Wiener-Hopf method. This function is different from zero only for $\eta <0$  (its Fourier image in analytic in the lower half-plane).
In the Fourier space, $X(\omega)$ can be read off from the results in \cite{Chen:2014vka} (we keep the original notation):
\begin{eqnarray} \label{Xfull}
X(\omega)&=&
 -\frac{\sqrt{i \pi } \pi  \left(M^2+1\right) \Gamma \left(\frac{i \omega }{2 \pi }\right)^2 e^{-\frac{i \omega  \phi (M)}{2 \pi }}}{2 \sqrt{\omega -i \epsilon }\: \Gamma \left(-\frac{(M-i) \omega }{2 \pi }\right) \Gamma \left(\frac{(M+i) \omega }{2 \pi }\right)}\times \\
         &\:& \times \sum_{n=1}^{\infty} \frac{(-1)^n}{n n!} \left(\frac{e^{-\frac{i n \phi (M)}{M+i}} \Gamma \left(\frac{(M-i) n}{M+i}\right)}{\left(\omega +\frac{2 \pi  n}{M+i}\right) \Gamma \left(-\frac{i n}{M+i}\right)^2}-\frac{e^{\frac{i n \phi (M)}{M-i}} \Gamma \left(\frac{(M+i) n}{M-i}\right)}{\left(-\omega
+\frac{2 \pi  n}{M-i}\right) \Gamma \left(\frac{i n}{M-i}\right)^2}\right) \nonumber
\end{eqnarray}
where
\begin{equation}
 \phi(M)=2 M \arctan(M)-\log \left(M^2+1\right).
\end{equation}

Let us discuss the analytical structure of this function. 
The sum has two branches of simple poles in the upper half complex plane of $\omega$, which comes from the $G_-(\omega)$ defined in \cite{Chen:2014vka}.
The prefactor, which is essentially $1/G_-(\omega)$, contains double poles, from $\Gamma \left(\frac{i \omega }{2 \pi }\right)^2$.
There is also a square root cut in the positive imaginary axis. 
Therefore, inverse Fourier transforming back to the coordinate space for this exact expression is very difficult. 
As we are interested to probe the phase structures at the decompactification limit, we will study this function at this regime, where the treatment simplifies much. 

\subsection{Universal regime}

To study $\eta =O(M)$, we deal with $\omega=O(M^{-1})$. At the decompactification limit $M \rightarrow \infty$, 
the sum part of \eqref{Xfull} reduces to (equation (4.6) in \cite{Chen:2014vka}):
\begin{equation}
 \text{sum}=\frac{1}{M^2 \omega }-\frac{\,{\rm e}\,^{\frac{1}{2} i M \omega }}{2 M \sin \left(\frac{M \omega }{2}\right)}
\end{equation}
and a naive large M approximation simplifies the prefactor to:
\begin{equation}
  \text{prefactor}=-\frac{(-1)^{1/4} \pi ^{3/2} M^3 e^{-\frac{1}{2} i M \omega } \sin \left(\frac{M \omega }{2}\right)}{\omega  \sqrt{\omega -i \epsilon }}
\end{equation}
Hence, the full expression is:
\begin{equation}
 X(\omega )=-\frac{(-1)^{1/4} \pi ^{3/2} M^2}{2 \omega  \sqrt{\omega -i \epsilon }} 
	  \left(-1+\frac{e^{-\frac{1}{2} i M \omega } \sin \left(\frac{M \omega }{2}\right)}{\frac{M \omega }{2}}\right)
	  \label{X_w_M}
\end{equation}
We see that $ X(\omega )$ does not have poles anymore. This is because the simple poles got canceled by the zeros of the prefactor.
The double poles are not relevant at this regime, as $\Gamma \left(\frac{i \omega }{2 \pi }\right)$ is approximated by its expansion at zero. 
Only the cut remains in the upper half complex plane. 

Straightforward contour integration gives us the expression in the coordinate space, which is:
\begin{equation}
 X(\eta )=-\frac{ \pi  M }{3}\left[2 (-M-\eta )^{3/2} \theta (-M-\eta )+\sqrt{-\eta }  (3 M+2 \eta ) \theta (-\eta )\right]
 \label{X_v}
\end{equation}
Upon rescaling, this is the same as eq.~(\ref{func-X}) in the main text. 

\subsection{Non-universal regime}

The solution \eqref{X_v} for $X(\eta )$ obtained in the regime $O(M)$ actually approximates well this regime too.

In Fourier space, following the simplifications explained in the appendix C of \cite{Chen:2014vka},
$X(\omega)$ reduces to: 
\begin{equation}
 X(\omega) =-\dfrac{(-1)^{1/4} \pi ^{3/2} M^2}{2 \omega  \sqrt{\omega -i \epsilon }}
	   \left(-1+\frac{i \omega }{4 \pi^2  M}
	    \Gamma \left(\frac{i \omega }{2 \pi }\right)^2  e^{-\frac{i \omega}{\pi } \left(-1+\log \frac{i \omega }{2 \pi }\right)} 
	    \right)
\label{X_w_1}
\end{equation}
Due to its complicated analytical structure, especially the double poles and the logarithmic branch cuts in the exponential, 
we abstain from computing its exact expression in coordinate space.
Instead, we study its asymptotic behaviors, for $\omega \gg 1$ and $\omega \ll 1$. 
For the former limit, the first term of the sum in \eqref{X_w_1} dominates, which is the same as the first term in \eqref{X_w_M} from the $O(M)$ regime. 
For $\omega \ll 1$, the leading contribution comes from the second term of the sum in \eqref{X_w_1}, 
and we expect matching with the regime $O(M)$, which it does. Here are the asymptotic results in the coordinate space:  
\begin{eqnarray}
 |\eta |<<1: \quad  X(v)&=& - M^2 \pi \eta   			\\
 |\eta |>>1: \quad X(v)&=&\dfrac{2 M \pi}{3} (-\eta )^{3/2} 	
\end{eqnarray}
Numerically, we also see that \eqref{X_v} approximates this regime very well at large M.

\bibliographystyle{nb}

\end{document}